\documentclass[12pt]{iopart}

\usepackage{graphicx}
\begin{document}

\title[]{Evidence for Coexistence of Superconductivity and Magnetism in Single Crystals of Co-doped SrFe$_2$As$_2$}

\author{Jun Sung Kim, Seunghyun Khim, Liqin Yan, N. Manivannan, Yong Liu, Ingyu Kim, G. R. Stewart$^{\dag}$, and Kee Hoon Kim$^{*}$}
\address{FPRD, Department of Physics and Astronomy, Seoul National University, Seoul 151-747, Republic of Korea}
\ead{khkim@phya.snu.ac.kr}

\begin{abstract}
In order to investigate whether magnetism and superconductivity
coexist in Co-doped SrFe$_2$As$_2$, we have prepared single crystals
of SrFe$_{2-x}$Co$_x$As$_2$, $x$ = 0 and 0.4, and characterized them
via X-ray diffraction, electrical resistivity in zero and applied
field up to 9 T as well as at ambient and applied pressure up to 1.6
GPa, and magnetic susceptibility.  At $x$ = 0.4, there is both
magnetic and resistive evidence for a spin density wave transition
at 120 K, while $T_c$ = 19.5 K - indicating coexistent magnetism and
superconductivity.  A discussion of how these results compare with
reported results, both in SrFe$_{2-x}$Co$_x$As$_2$ and in other
doped 122 compounds, is given.
\end{abstract}

\maketitle

\section{Introduction}

The recent discoveries of ever-mounting transition temperatures in
the superconducting iron arsenside 2-dimensional layered compounds,
coupled with the goal of understanding the pairing mechanism(s) of
this newly discovered class of superconducting compounds, has lead
to a surge of activity in materials-based condensed matter physics.
>From a superconducting transition temperature $T_c$ = 26 K in
LaFeAs(O$_{1-x}$F$_x$)\cite{1} the value is now up to $T_c$ = 55 K
in SmFeAs(O$_{1-x}$F$_x$)\cite{2}. Of particular help in the quest
for understanding this new physics has been the widening range of
compounds in which the "iron arsenide (FeAs)" based
superconductivity has been found, moving from the rather difficult
materials synthesis of the original 1111 compounds with F-doping to
the more-easily-prepared 122 compounds (non-superconducting
prototype BaFe$_2$As$_2$) discovered by Rotter \emph{et al.}\cite{3}
These latter compounds, as was pointed out by Ni \emph{et
al.}\cite{4} can be rather easily grown from a Sn flux as well as
from an FeAs 'self flux'\cite{5}. Thus, much of the recent effort
for elucidating the physics has focused on these 122 compounds, with
both polycrystalline and single crystal work.  Single crystals of
course allow greater homogeneity and the possibility of following
the anisotropy of the fundamental properties - often important in
distinguishing the underlying mechanisms of
superconductivity\cite{6}.

A central question\cite{7} for deciding on the superconducting
pairing mechanism in these FeAs superconductors has been the
interplay/relationship between the ubiquitous magnetic behavior in
the undoped, non-superconducting starting compounds (either the 1111
family or the $A$Fe2As2, where $A$ = Ca, Sr, Ba, and Eu) which is
then suppressed by the doping.  In the 122 family, K/Na/Cs, or hole
doping, on the $A$-site or Co/Ni - electron doping\cite{5} - on the
Fe site) induces superconductivity. Whether the magnetic (spin
density wave, 'SDW') behavior is coupled to the occurrence of
superconductivity in SrFe$_2$As$_2$ doped with Co is a main subject
of the present work, using single crystals prepared in Sn flux.

The question "does the SDW coexist with superconductivity in FeAs
superconductors?" might seem straightforward to answer. However,
even in just the 122 compounds, there exist at present four starting
compounds $A$Fe$_2$As$_2$ ($A$ = Ca, Sr, Ba, and Eu) with both hole
(including work on Na, K and Cs) and electron (Co and Ni) doping,
and as well the very important materials aspects of both single- and
poly-crystalline samples.  Even a cursory review of the current
status of this 4 (Ca, Sr, Ba, Eu) $\times$ 2 (hole/electron)
$\times$ 2 (single/poly) 'phase space' for just the 122 compounds
already reveals both large differences but also serious conflicts
between the various results.  The rate at which doping depresses
$T_{SDW}$ and induces superconductivity varies widely between the
various $A$ atoms and either hole- or electron-dopants, which is a
sign of the richness of this new class of materials.  However, there
are also conflicts in some results on the same $A$ atom and the same
dopant which involve disagreements in concentration dependence of,
\emph{e. g.}, $T_{SDW}$, in whether the SDW transition is first or
second order in, \emph{e. g.}, SrFe$_2$As$_2$\cite{8,9,10,11}, and
even in the quite fundamental question of coexistence of magnetism
and superconductivity itself (see Table 1). Our work on the electron
doped SrFe$_2$As$_2$ is the first to be done on single crystals in
this compound (with one report on polycrystalline samples\cite{12}
and one on thin films\cite{13}), bringing an initial data set for
the 4 $\times$ 2 $\times$ 2 set closer to completion. SrFe$_2$As$_2$
single crystals show a structural phase transition from a
high-temperature tetragonal phase to a low-temperature orthorhombic
phase at the same temperature as the SDW, $T_o$ = 198 K \cite{11}
similar to the behavior observed in the BaFe$_2$As$_2$
compound\cite{4}.

As summarized in Table 1, the relation between magnetic behavior and
superconductivity in the 122 FeAs superconductors has been addressed
quite thoroughly for $A$ = Ba, but somewhat less so for $A$ = Ca,
Sr, and Eu.  There is also growing work on electron doping
(primarily Co replacing Fe) for all the $A$ species listed. As
detailed in Table 1, at present the question of whether magnetism in
the form of a SDW coexists with superconductivity in doped
$A$Fe$_2$As$_2$ is still controversial.

Some of the disagreement in resolving the issue of coexistence of
magnetism and superconductivity in the doped 122 $A$Fe$_2$As$_2$
materials made apparent by the summary in Table 1 can be resolved as
merely based on interpretation.  For example, some authors (\emph{e.
g.} see Refs. \cite{17,18}) have stated that the SDW transition is
suppressed based on the lack of sharp structure in $\rho$ vs $T$
data, although a shoulder that might be indicative of a weak
transition exists in their data. However, some of the disagreements
appear to be fundamentally unresolvable at this time.  One example
of this involves contrasting $T_{SDW}$ vs $x$ results even in high
quality single crystals of BaFe$_{2-x}$Co$_x$As$_2$ by X. F. Wang
\emph{et al.}\cite{21} and by J.-H. Chu \emph{et al.}\cite{22} Such
disagreement is independent of any interpretation.

Two important lessons to be drawn from the summary in Table 1 on
single crystal SrFe$_{2-x}$Co$_x$As$_2$ are:  (1) a fine gradation
in composition in BaFe$_{2-x}$Co$_x$As$_2$ was shown to be necessary
for determining whether $T_{SDW}$ has been suppressed to $T$ = 0
when superconductivity first appears\cite{21,22,23} (2) Some of the
work on polycrystalline samples has been found to disagree with
single crystal work, partly at least for reasons still under
discussion, thus obscuring any possible conclusions. In general,
although single crystals grown in Sn-flux can have small inclusions
of Sn\cite{4}, single crystals should be more homogeneous than
sintered polycrystalline material. In order to address point (1), we
are working on single crystals of SrFe$_{2-x}$Co$_x$As$_2$, $x$ =
0.1, 0.2, 0.3, and 0.5 in addition to the work on $x$ = 0 and 0.4
reported here. However, as will be discussed below, $T_{SDW}$ is
suppressed much less rapidly with Co in SrFe$_{2-x}$Co$_x$As$_2$
than in BaFe$_{2-x}$Co$_x$As$_2$, and our present work on $x$ = 0
and 0.4 is sufficient to show the coexistence of magnetism and
superconductivity in SrFe$_{2-x}$Co$_x$As$_2$ - contradicting
conclusions based on polycrystalline
SrFe$_{2-x}$Co$_x$As$_2$\cite{12}(see Table 1).

\begin{table}
\caption{\label{tabone}Survey of previous doping results in 122 FeAs
superconductors. Units of temperature are Kelvin; results are for
either single- or poly-crystalline samples. It is worth noting that
some authors, well-focused on the difficulty of answering the
coexistence question precisely, have used more precise determination
of $T_{SDW}$ (\emph{e. g.}, Wang \emph{et al.}\cite{21} used
specific heat; Zhang \emph{et al.}\cite{16} used band splitting
measured by photoemission). }

\begin{indented}
\lineup
\item[]\begin{tabular}{@{}*{7}{l}}
\br $A_{1-x}A'_{x}$Fe$_{2-y}$Co$_y$As$_2$ &Dopant$_{x,y}$ &$T_{SDW}$
&$T_c$ &Coexistence & single &[Ref.]\\\ns ($A'$ = K,
Na)&&&&(yes/no)&/poly&\cr \mr

$A$ = Ca &Co$_{0.06}$&none&17&no&single&[14]\cr
&Na$_{0.5}$&none&20&no&poly&[15]\cr\mr

$A$ = Sr & K$_{0.2}$& 135 & 25 & yes & single &[16]\cr
&K$_{0.4}$&none&38&no&single&[17]\cr

&K$_{0.4}$&none&20$^*$&no&poly&[18]\cr

&Na$_{0.5}$&160&35&yes&single&[19]\cr

&Co$_{0.2}$&none&19.2&no&poly&[12]\cr\mr

$A$ = Ba &K$_{0.5}$&70&37&yes&single&[19]\cr

&K$_{0.2,0.3}$&120,100&7,14&yes&poly&[20]\cr

&Co$_{0.17}$&75&9&yes&single&[21] \cr

&Co$_{0.10}$&35-50(split)&20&yes&single&[22]\cr

&Co$_{0.16}$&none&22&no&single&[5,23]\cr\mr

$A$ = Eu &K$_{0.5}$&none&32&no&poly&[24]\cr

&Pressure&115&30&yes&single&[25]\cr \br
\end{tabular}
\item[] $^*$ Annealed polycrystalline Sr$_{0.6}$K$_{0.4}$Fe$_2$As$_2$ changes $T_c$ from 38 to 20 K. In the unannealed state, there is an anomaly in $\rho$ at 200 K indicative
of SDW and $T_c$ = 38 K \cite{18}.
\end{indented}
\end{table}

\section{Experimental}

Single crystals of Co-doped SrFe$_2$As$_2$ were grown using high
temperature solution growth techniques with a Sn flux\cite{4}.
Stoichiometric amounts of the elemental Sr, Fe, Co and As were added
to Sn with the ratio of [SrFe$_{2-x}$Co$_x$As$_2$] : Sn = 1 : 20 and
placed in an alumina crucible, which was sealed in a silica ampoule
in vacuum. All the handling of the elements was performed in a glove
box with an Ar atmosphere (oxygen $<$ 1 ppm, H$_2$O $<$ 1 ppm). The
sealed crucible was heated to 700 $^{\rm o}$C (duration of 4 hours),
then to 1100 $^{\rm o}$C (duration of 4 hours). After this, the
sample was slowly cooled down to 500 $^{\rm o}$C at the rate of 4
$^{\rm o}$C /hour and then the plate-like single crystals of typical
dimensions 10 $\times$ 10 $\times$ 0.5 mm$^3$ were removed from the
Sn flux by centrifuging\cite{4}.

Resistivity measurements were made by a standard 4-wire ac method,
using a Quantum Design PPMS$^{\rm TM}$  system in fields up to 9 T.
Due to the large flat faces of the crystals, where the c-axis is
perpendicular to the face, alignment of the field either parallel to
the c-axis or in the ab-plane was straightforward.  Magnetic
susceptibility measurements were performed in the same Quantum
Design PPMS$^{\rm TM}$ system.

\section{Results and discussion}

\begin{figure}
\includegraphics[width=10cm,angle=0,bb=28 73 560 760]{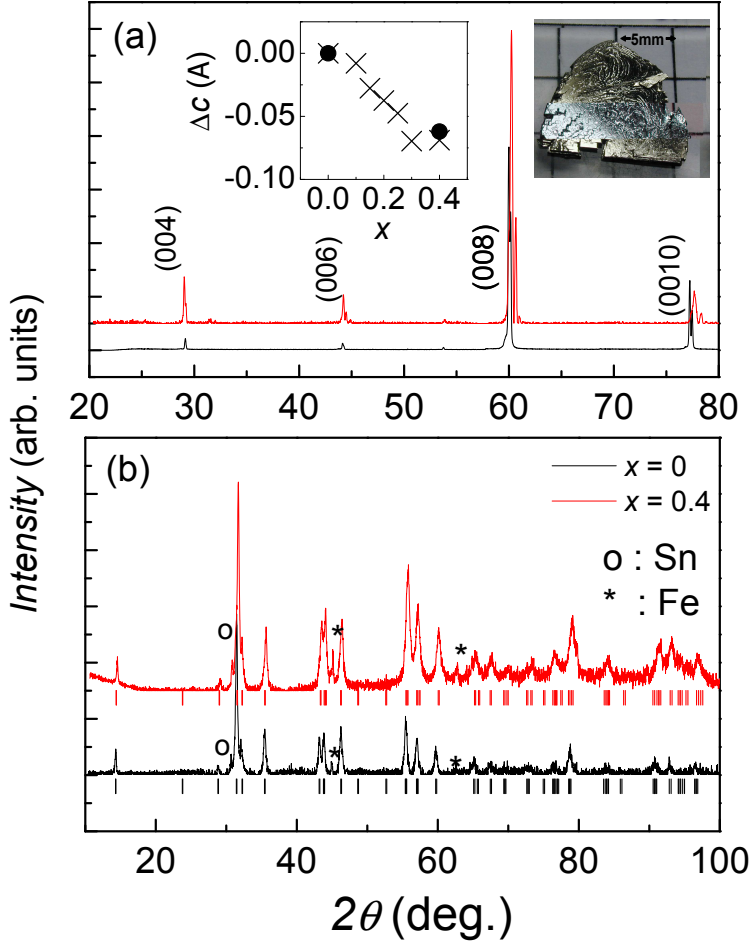}
\caption{(Color online) X-ray diffraction patterns for (a) single
crystal and (b) crushed powder of SrFe$_{2-x}$Co$_x$As$_2$ with $x$
= 0 and 0.4.  The upper left inset shows the decrease of the
$c$-axis lattice constant ($\Delta c$) in the single crystals of the
present work due to Co doping (solid circle). For comparison, we
also plot the $\Delta c$ vs $x$ for polycrystalline
SrFe$_{2-x}$Co$_x$As$_2$ ($\times$ symbols)\cite{12}. The upper
middle inset shows a photo of a $x$ = 0.4 single crystal. Note the
marked second phase lines in the powder pattern (b), where in
addition to a slight amount of Sn inclusion from the flux, some
excess Fe is seen.}
\end{figure}

X-ray diffraction measurements were carried out on a single crystal
from both of the compositions $x$ = 0 and 0.4. As shown in Fig. 1,
only (00$l$) reflections with even $l$ appear, indicating that the
$c$-axis is perpendicular to the crystal plate. The addition of Co
decreased the $c$-axis lattice parameter in these single crystals.
However, for reasons that will become clear below in the discussion
of the resistivity data for $x$ = 0.4, in order to investigate
possible crystal inhomogeneity and impurities below the rather
shallow penetration of the X-ray beam ($\sim$ a few $\mu$m) in the
single crystals, we undertook X-ray diffraction of powders made of
individual single crystals.  These data, shown also in Fig. 1,
provide a measurement of both the $a$- and $c$-axis lattice
parameters and are more characteristic of the bulk of the crystal.
These powder diffraction lattice parameter results agree with the
single crystal results. The results for $x$ = 0 were $a$ = $b$ =
3.928(3) ${\rm \AA}$, $c$ = 12.392(9)$\rm {\AA}$, while $a$ = $b$ =
3.925(3) ${\rm \AA}$, $c$ = 12.33(1) ${\rm \AA}$ for $x$ = 0.4. For
$x$ = 0, the $c$-axis lattice parameter is consistent with some
previous reports on poly-\cite{8} and single-crystal
materials\cite{17}, but is slightly larger than the polycrystalline
results of Leithe-Jasper \emph{et al.}\cite{12} With Co doping, it
has been shown that the c-axis lattice parameter decreases linearly
with Co concentration\cite{12,21,22}.

Considering the inconsistency in the the absolute value of the
$c$-axis lattice parameters in the literature, we focused here on a
comparison of the \emph{change}(contraction) of the $c$-axis in our
Co-doped crystals with that found in polycrystalline works\cite{12}
(see Fig. 1). According to Leithe-Jasper \emph{et al.} a $c$-axis
contraction of -0.07(1)${\rm \AA}$ is expected for $x$ = 0.4, which
is comparable with our result of -0.06(2) ${\rm \AA}$ for our
Co-doped single crystal. This provides a bulk proof for the presence
of approximately $x$ = 0.4 Co in our Co-doped single crystals.

\begin{figure}
\includegraphics[width=10cm,angle=0,bb=10 10 240 300]{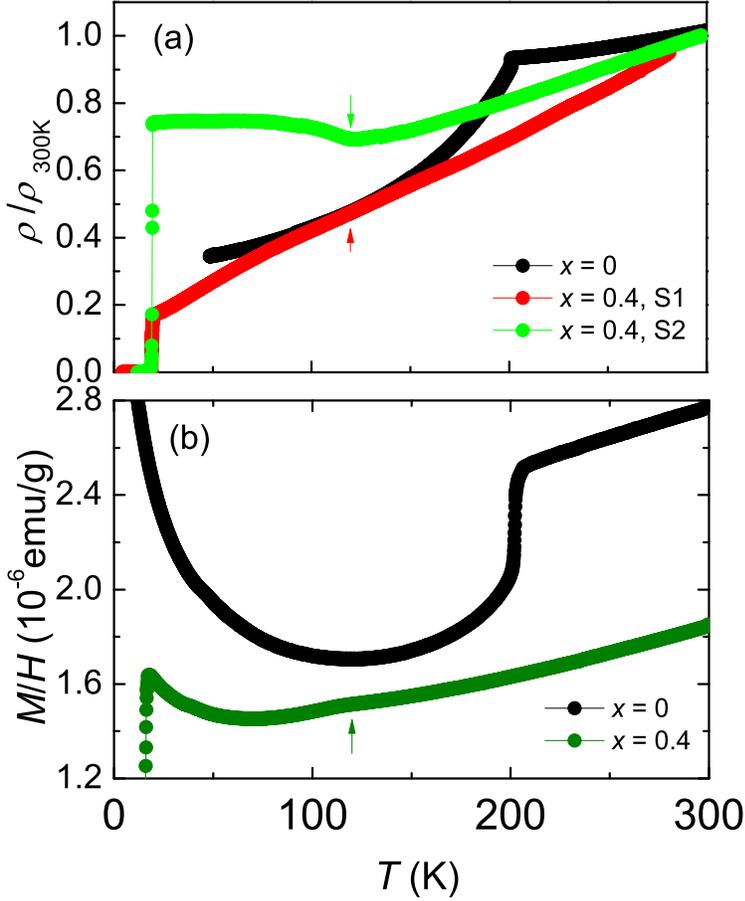}
\caption{(Color online) Resistivity vs temperature (a) for single
crystal SrFe$_{2-x}$Co$_x$As$_2$, $x$ = 0 and 0.4 (two samples from
the same growth batch, labeled S1 and S2), showing the anomalies at
$T_{SDW}$ (202 K for $x$ = 0 and $\approx$ 120 K for $x$ = 0.4). On
an expanded plot, not shown, extrapolations of the higher
temperature and separately the lower temperature resistivity data
from the data for S1 intersect at the temperature marked by the red
arrow. The plot in (b) shows the $M/H$ (measured in 7 T), where the
SDW anomalies for both $x$ = 0 and 0.4 are clear.}
\end{figure}

Resistivity and susceptibility data for $x$ = 0 and 0.4 are shown in
Figure 2.  Discussing the normal state properties first, as shown
clearly in Fig. 2, our single crystals for $x$ = 0.4 show differing
resistivity behaviors below $T_{SDW}$:  one crystal (S2) shows
evidence for strong scattering below $T_{SDW}$ while another crystal
(S1) shows only a slight change in slope (marked by the red arrow).
[We have measured a total of 6 different single crystals out of the
same growth batch for $x$ = 0.4, and the strong increase in $\rho$
below $T_{SDW}$ is found in two samples.  We are continuing to
investigate this.]  This sample dependence is of course reminiscent
of early sample dependence problems in $\rho$ in
YBa$_2$Cu$_3$O$_{7-\delta}$. However, in both crystals (as well in
all the other crystals measured from this $x$ = 0.4 batch), the
superconducting transition is consistently at $T_c$ = 19.5 K.
Clearly, the magnetic anomaly for SrFe$_{1.6}$Co$_{0.4}$As$_2$ is a
clearer evidence for a SDW transition at 120 K than the slight
change of slope/broad hump in the resistivity data that is
characteristic of most of our samples. In the polycrystalline work
on SrFe$_{2-x}$Co$_x$As$_2$\cite{12}, the resistivity curve for a
nonsuperconducting sample of $x$ = 0.1 increases below $T$ $\sim$
130 K similar to the S2 data for the $x$ = 0.4 single crystal in
Fig. 2. The polycrystalline resistivity data\cite{12} for $x$ $\geq$
0.2 ($T_c$ = 19.2 K for $x$ = 0.2)  show positive curvature vs
temperature between $T_c$ and 300 K, \emph{i.e.} unlike both the S2
and S1 resistivity curves for our single crystal
SrFe$_{2-x}$Co$_x$As$_2$ shown in Fig. 2. Thus, if it were not for
the good agreement in the lattice contraction for the same
compositions ($x$ = 0 and 0.4) in the present single crystal work
compared to the polycrystalline work\cite{12}, both the difference
in the behavior of $T_{SDW}$ and $T_c$ would have called the
comparability of the Co-compositions into question. As it is, it
would be useful for magnetic susceptibility data to higher
temperatures than 25 K (the upper limit in Ref. 12) to be measured
on the polycrystalline samples.  At this time it is not clear why
there is disagreement between $T_{SDW}$($x$) results determined by
resistivity data on poly-\cite{12} and single-crystalline (present
work) samples of SrFe$_{2-x}$Co$_x$As$_2$.

As stated in the Introduction, the field of FeAs superconductivity
is in a state of flux at present. The variation of the resistivity
seen in our single crystals for $x$ = 0.4, and the disagreement
between our single crystal compositional dependence of $T_{SDW}$ and
$T_c$ compared to polycrystalline\cite{12} results is perhaps one
reason why some of these open questions must remain open until
better understanding of sample quality is achieved.

The superconducting transition temperature for $x$ = 0.4
SrFe$_{2-x}$Co$_x$As$_2$ single crystals is 19.5 K, which is
comparable with the maximum $T_c$ achieved by Co-doping in
polycrystalline SrFe$_2$As$_2$\cite{12} and coexists with the
magnetic transition at $T_{SDW}$ $\approx$ 120 K.  Our work in
progress on other compositions confirms this result, adding one more
piece to the conclusion that is becoming clearer (see Table 1 and
references therein) that - contrary to early conclusions, magnetism
and superconductivity clearly coexist in these 122 FeAs
superconductors. In comparison with Co-doped BaFe$_2$As$_2$, the
magnetic phase is more robust in Co-doped SrFe$_2$As$_2$. For
BaFe$_2$As$_2$, the $T_{SDW}$ is decreased rapidly with a relatively
small amount of Co substitution, \emph{i. e.}, $x$ = 0.12, which is
sufficient to fully suppress the SDW transition and induce the
maximum $T_c$ $\approx$ 24 K. In contrast, we still observe the
clear magnetic transition at $T_{SDW}$ $\approx$ 120 K with $x$ =
0.4 in SrFe$_{2-x}$Co$_x$As$_2$ with $T_c$ $\approx$ 20 K. This
result may be related to the higher $T_{SDW}$ $\approx$ 202 K in
SrFe$_2$As$_2$ than that of BaFe$_2$As$_2$ ($T_{SDW}$ $\approx$ 140
K).

\begin{figure}
\includegraphics[width=10cm,angle=0,bb=10 10 222 235]{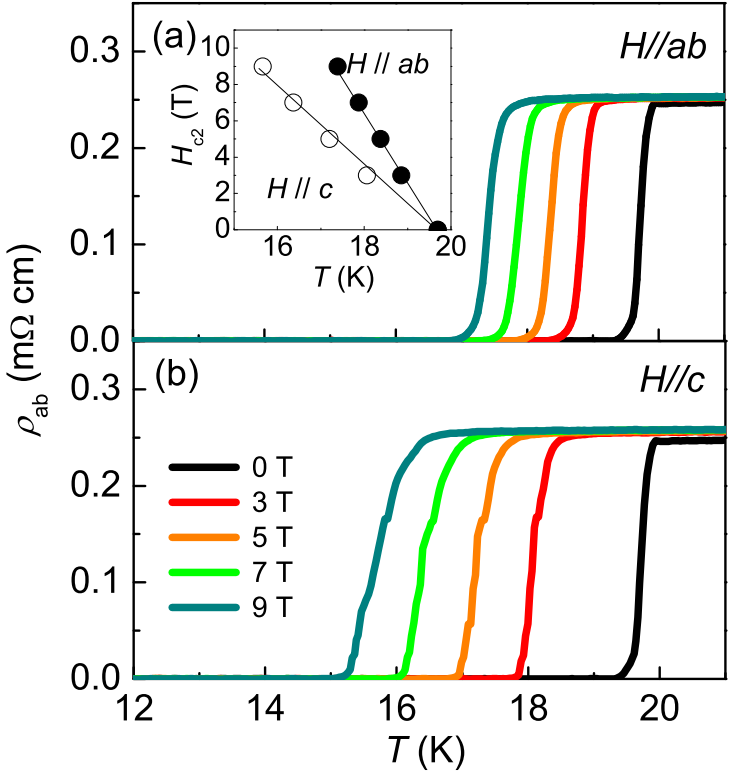}
\caption{(Color online) Temperature dependence of the $ab$-plane
resistivity of single crystal SrFe$_{1.6}$Co$_{0.4}$As$_2$ with
different magnetic fields along the $ab$-plane (a) and the $c$-axis
(b). The inset shows the $H_{c2}$($T$) curves near $T_c$ for the two
field directions, $H$ $\parallel$ $c$ and $H$ $\parallel$ $ab$. }
\end{figure}

The temperature-dependence of $H_{c2}$($T$), defined by 90\% of the
resistive transition is shown in the inset of Fig. 3. Both
$H_{c2}^{ab}$ and $H_{c2}^{c}$ showed almost linear temperature
dependence with slopes of $dH_{c2}^{ab}/dT$ = -3.9 T/K and
$dH_{c2}^{c}/dT$ = -2.2 T/K.  The zero temperature upper critical
fields can be estimated using the Werthamer-Helfand-Hohenberg
formula, $H_{c2}$(0) = -0.69$T_c$($dH_{c2}/dT$)$|$$_{T_c}$, yielding
$H_{c2}^c$(0) = 30 T and $H_{c2}^{ab}$(0) = 53 T. The corresponding
coherence lengths are 33 ${\rm \AA}$ and 19 ${\rm \AA}$ along the
$ab$-plane and the $c$-axis, respectively. The $c$-axis coherence
length is comparable with the distance between two adjacent FeAs
layers, $d$ $\sim$ 6 ${\rm \AA}$, indicating the quasi-2
dimensionality of the superconductivity. The anisotropy parameter
$\Gamma$ = $H_{c2}^{ab}/H_{c2}^c$ derived from the data in Fig. 3 is
$\Gamma$ $\approx$ 1.7, which is comparable with $\Gamma$ $\approx$
1.5-2 of K- or Co-doped BaFe$_2$As$_2$ and K-doped SrFe$_2$As$_2$
\cite{17} but significantly lower than $\Gamma$ $\approx$ 5-10 in
the 1111 oxypnictides.

\begin{figure}
\includegraphics[width=10cm,angle=0,bb=10 10 255 230]{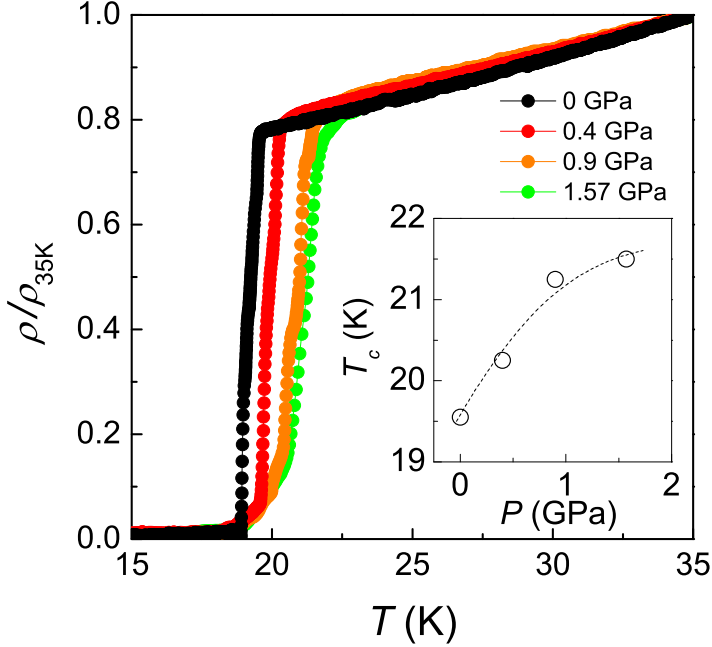}
\caption{(Color online) Superconducting transition temperatures for
single crystal SrFe$_{1.6}$Co$_{0.4}$As$_2$ determined by the
resistivity as a function of hydrostatic pressure}
\end{figure}

As a final characterization of the superconductivity we observe in
our single crystals of SrFe$_{1.6}$Co$_{0.4}$As$_2$, we present the
pressure dependence of $T_c$ in Fig. 4. Gooch \emph{et al.}\cite{26}
reported on $T_c$($P$) in polycrystalline
Sr$_{0.6}$K$_{0.4}$Fe$_2$As$_2$, $T_c^{onset}$ = 37 K, and find an
increase in $T_c^{onset}$ at 0.9 GPa of about 1.2 K, or about 3 \%,
compared to our result for electron-doped SrFe$_2$As$_2$ where $T_c$
increases by about 1.8 K, or about 9 \% with 0.9 GPa. Gooch \emph{et
al.} also see some saturation in the rise of $T_c$ with pressure in
their 1.7 GPa data comparable to what we observe (see inset to Fig.
4). From previous pressure work on the K-doped $A$Fe$_2$As$_2$ ($A$
= Ba, Sr) compounds, it has been found that the pressure dependence
of $T_c$ reflects the "dome" shape of the doping dependence of
$T_c$\cite{26}. The underdoped and overdoped samples show positive
and negative pressure dependence, respectively, while almost no
pressure dependence of $T_c$ is observed in the optimally doped
sample. As mentioned already, our $x$ = 0.4 crystal shows $T_c$ =
19.5 K, close to the maximum $T_c$ of Co-doped polycrystalline
SrFe$_2$As$_2$\cite{12}, thus in the optimal doping regime. The
sizable pressure dependence of $T_c$ in our crystal, therefore,
suggests that there is still room for improving the superconducting
transition temperature by further tuning, \emph{e. g.}, using
external pressure. Similar behavior has been also observed in
optimally Co-doped BaFe$_2$As$_2$\cite{27}. This different behavior
between K-doped and Co-doped 122 compounds indicates that the
pressure dependence of $T_c$ is determined not just by the doping
level of the FeAs layer but also reflects more complex interplay
with other parameters such as the degree of hybridization between
the Fe and As states that can be tuned by, \emph{e. g.}, a bonding
angle of the Fe-As-Fe network\cite{28}.

\section{Summary and Conclusion}

Our present work on single crystal SrFe$_{2-x}$Co$_x$As$_2$ ($x$ = 0
and 0.4) shows clear signatures in both electrical resistivity and
magnetization curves for the presence of a spin density wave at 202
and 120 K, respectively.  The $x$ = 0.4 sample shows
superconductivity at 19.5 K, which - in the spirit of the work on
the FeAs superconductors to date (see Table 1) - allows the
conclusion that superconductivity is coexistent with magnetism (SDW)
in single crystal SrFe$_{1.6}$Co$_{0.4}$As$_2$. Of course, a
microscopic determination of the coexistence below $T_c$ is further
required. Both the single crystal and powder X-ray diffraction
characterization of our samples show internal consistency as well as
agreement of the lattice contraction with Co doping, compared to the
polycrystalline work on SrFe$_{2-x}$Co$_x$As$_2$\cite{12}. In
contrast, our compositional dependence of both $T_{SDW}$ and $T_c$
disagree with the polycrystalline data in Ref. 12 which does not
report magnetic susceptibility. The anisotropy of the upper critical
field $H_{c2}$ in our single crystals of
SrFe$_{1.6}$Co$_{0.4}$As$_2$ is consistent with K- or Co-doped
BaFe$_2$As$_2$ and K-doped SrFe$_2$As$_2$\cite{17}. The pressure
dependence of $T_c$ of our single crystalline
SrFe$_{1.6}$Co$_{0.4}$As$_2$ is, when expressed as a percentage of
$T_c$($P$ = 0), much larger than that observed\cite{26} in K-doped
SrFe$_2$As$_2$.

An important conclusion that can be drawn from our present work is
that even in single crystals there appears to be significant sample
dependence at least in the resistivity below $T_{SDW}$, while
$T_{SDW}$ and $T_c$ themselves did not show any sample dependence.
Our results clearly show sample dependence in the resistivity, as
well as an unexplained difference between our single crystal and
Ref. 12's polycrystalline values of $T_{SDW}$ and $T_c$ as a
function of Co-concentration. This may be a useful cautionary note
about the rush to make firm conclusions in the early stages of the
fascinating study of superconductivity in the 122 FeAs compounds.

\ack
This work was supported by the National Research Lab program
(M10600000238) and KICOS through a grant provided by the MEST
(K20702020014-07E0200-01410).  Work at Florida supported by the US
Department of Energy, contract no. DE-FG02-86ER45268.  JSK is
supported by BK21 Frontier Physics Research Division.

\section*{References}

$^*$Corresponding author.\\
$^{\dag}$On sabbatical from Department of Physics, University of
Florida.

\end{document}